\documentclass[12pt,preprint]{aastex}

\shorttitle{A power-law break in the PSD of Sgr~A*}
\shortauthors{Meyer et al.}

\begin{document}


\title{A power-law break in the near-infrared power spectrum \\
    of the Galactic center black hole}


\author{L. Meyer, T. Do, A. Ghez, M. R. Morris, S. Yelda}
\affil{Department of Physics and Astronomy, University of California,
    Los Angeles, CA 90095-1547}

\author{R. Sch\"odel}
\affil{Instituto de Astrof\'{i}sica de Andaluc\'{i}a -- CSIC, Camino Bajo de Hu\'{e}tor 50, 18008 Granada, Spain}

\author{A. Eckart}
\affil{Universit\"at zu K\"oln, Z\"ulpicher Str. 77, 50937 K\"oln, Germany}



\begin{abstract}
Proposed scaling relations of a characteristic timescale in the X-ray power spectral density of galactic and supermassive black holes have been used to argue that the accretion process is the same for small and large black holes. Here, we report on the discovery of this timescale in the near-infrared radiation of Sgr~A*, the $4\cdot 10^6\,\mbox{M}_\sun$ black hole at the center of our Galaxy, which is the most extreme sub-Eddington source accessible to observations. Previous simultaneous monitoring campaigns established a correspondence between the X-ray and near-infrared regime and thus the variability timescales are likely identical for the two wavelengths. We combined Keck and VLT data sets to achieve the necessary dense temporal coverage, and a time baseline of four years allows for a broad temporal frequency range. Comparison with Monte Carlo simulations is used to account for the irregular sampling. We find a timescale at $154^{+124}_{-87}$\,min (errors mark the 90\% confidence limits) which is inconsistent with a recently proposed scaling relation that uses bolometric luminosity and black hole mass as parameters. However, our result fits the expected value if only linear scaling with black hole mass is assumed. We suggest that the luminosity-mass-timescale relation  applies only to black hole systems in the soft state. In the hard state, which is characterized by lower luminosities and accretion rates, there is just linear mass scaling, linking Sgr A* to hard state stellar mass black holes.
 
\end{abstract}





\section{Introduction}

Cosmic black holes (BH) show a very wide range of masses: from stellar masses to hundreds of millions of solar masses. An open question has long been whether the accretion and variability processes occurring in the immediate vicinity of the event horizon are the same over this broad mass range. To test this intriguing possibility, investigators identified and studied a characteristic timescale associated with the aperiodic X-ray variability of black hole X-ray binaries (BHXRBs) and active galactic nuclei  \citep[AGN; see, e.g.,][]{uttley,markowitz03,uttley05}. This timescale corresponds to a break in the power spectral density (PSD) at a certain temporal frequency where a power-law of slope $\gamma$ (with $P(f) \propto f^{-\gamma}$) breaks to a steeper slope $\beta > \gamma$.

\citet{mchardy} proposed a scaling relationship between the break frequency, the mass of the black hole and the bolometric luminosity of its accretion flow. This extended earlier work which hypothesized that the break timescales of AGN scale linearly with BH mass from the timescales observed in BHXRBs, albeit with some scatter. \citet{mchardy}  found that this scatter can be explained by introducing the bolometric luminosity as a correction factor.  This lead them to conclude that AGN are scaled-up galactic BHs. However, their sample consists of only 10 AGN and it is therefore desirable to test their scaling relation with newly determined AGN break frequencies. The BH in our own Galactic center (identified with the radio source Sgr~A*) is an an especially interesting object, as it is the most underluminous BH accretion system observed thus far (with a bolometric luminosity nine orders of magnitude lower than its Eddington luminosity). It can therefore test the \citet{mchardy} relation in so far inaccessible regions of the parameter space. 

The existence of a supermassive BH in the center of the Milky Way has been demonstrated beyond reasonable doubt by the proper motion of stars detected in the near-infrared (NIR) waveband  \citep{eckigenzel, ghez98, gheznature,ghez,ghez08,genzel00,rainer1,gillessen08}. X-ray and NIR emission associated with Sgr~A* has been observable since 2001 and 2003, respectively \citep{baganoff01,genzel,ghez04}, which showed that it is highly variable at both wavelengths \citep[e.g.][]{ecki1,ecki2,ich2,ich,trippe}. 
Simultaneous NIR and X-ray monitoring campaigns revealed that each X-ray flare is accompanied by a NIR flare with zero time lag, which leads to the conclusion that the X-ray photons are being produced by Compton scattering off of the relativistic electrons which radiate in the NIR \citep{ecki04,ecki1,ecki08,danmultwav,yusef06,hornstein07}. The fact that the same bunch of electrons is the source for both the NIR and X-ray flux makes it possible to interpret the NIR timing studies reported here in the context of the AGN and BHXRB X-ray results. In fact, since the X-ray background is very high due to the lack of resolving power of X-ray telescopes  \citep[and thus only during the brightest flares is Sgr~A*  actually observed; see, e.g.,][]{baganoff01,belanger05}, a detailed study of the PSD is more complicated in the X-rays than in the NIR\footnote{We want to note that also in the NIR there might be a background due to unresolved stellar sources at the position of Sgr~A*, which has been estimated to contribute up to $30\%$ to the observed flux when Sgr~A* is at its lowest flux levels \citep{tuan}. This low background level is negligible for the purpose of our study.}.

Recent studies of the NIR properties of Sgr~A* showed that the variability on timescales of minutes to hours is completely described by an (unbroken) power-law PSD with a slope of -1.6 to -2.5 \citep[][see also Fig.~18 in Eckart et al. 2006a.]{ich08,tuan}.  In this \textit{Letter}, we extend the time baseline up to years by combining Keck and VLT data. We report for the first time the existence of a power-law break frequency in the NIR PSD of Sgr~A*, and we show that its deduced range is inconsistent with the \citet{mchardy} scaling relation, but fits the expected value when linear scaling with mass is assumed.


\section{The data}

The NIR adaptive optics instruments at Keck II (NIRC2) and at the VLT UT4 (NACO) have been used for Sgr~A* observations since 2003. A time baseline of years and a sampling timescale of minutes are needed to analyze the broadband PSD of Sgr~A* from minutes to years. Obviously, large gaps in the sampling pattern are unavoidable as the NIR observations have to be carried out during the night and available telescope time limits realistic data sets to a couple of nights each year. 

\begin{figure}
\includegraphics[scale=.6]{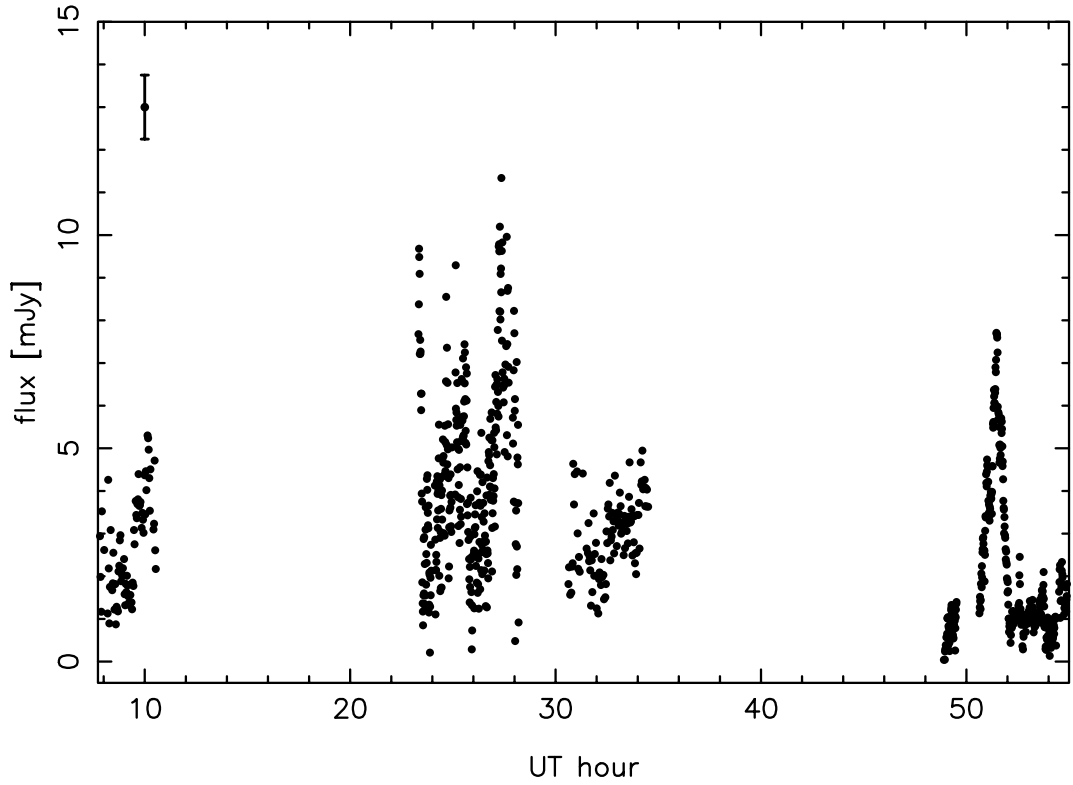}
\caption{The de-reddened flux of Sgr~A* on 2004 July 06 - 08. The abscissa shows the hours elapsed since July 06 00:00 UT. Error bars are omitted for clarity;  a sample is shown in the upper left corner. It is $\pm 0.75$\,mJy as determined from the structure function. The two big gaps mark the time of daylight.}
\label{lc}
\end{figure}

To cover the high-frequency to mid-frequency part of the PSD (meaning timescales of minutes to days in this context), we looked for Keck and VLT data from consecutive nights with individual night observations that are as long as possible. We identified the nights from 2004 July 06 -- 08 (all times UT) as the most suitable. Beginning on July 06, Sgr~A* was observed from 07:50 to 10:35 UT with Keck II, from 23:19 to 04:16 (July 07) with the VLT, then again from 06:35 to 10:30 with Keck II, and finally from 00:53 (July 08) to 06:53 with the VLT \citep[the Keck and VLT data have been published in][respectively]{hornsteinphd, ecki1}. Clearly, this dense coverage is only possible by combining the Keck and VLT data sets. Please note that these alternating observations were coincidental and not arranged. As the Keck data for these two nights were taken at L' (centered at $3.8\,\mu m$) and all other data were taken at K-band (centered on $2.15\,\mu m$)\footnote{More precisely, Keck uses a K' filter ($\lambda=2.12\,\mu m$) and VLT a K$_S$ filter ($\lambda=2.18\,\mu m$). We calibrated both data sets in the same way and thus the difference is negligible.}, we adopt the finding of \citet{hornstein07} that the NIR flux is consistent with a constant spectral index of $\alpha = -0.6$ ($F_\nu \propto \nu^\alpha$). Thus, we scaled the flux at L' down to the K-band level with this relation. The resulting lightcurve for the 3 nights is shown in Fig.~\ref{lc}. All fluxes are de-reddened using $A_K = 3.2$ as it was done in \citet{hornstein07}. Other authors have claimed variable and/or somewhat steeper spectral indices \citep[e.g.][]{gillessen06}. However, our results presented here are not very sensitive to the exact value of $\alpha$.

For the coverage of the mid-frequency to low-frequency part of the temporal spectrum (timescales of days to years) we averaged the data of individual nights. This leads to a lightcurve with irregular sampling of down to one day that covers a time baseline of 4 years. The details of the whole data set are given in Table~1. 

The data reduction was standard, i.e. flat fielding, sky subtraction, and correction for bad/hot pixels. The VLT data have been deconvolved with point-spread functions (PSF) extracted from the individual images \citep{diolaiti}. Aperture photometry was done on each image and the flux was calibrated relative to sources in the field with known flux. See \citet{ich08} for more details. For the Keck data the individual PSFs have been used to fit sources in the field; see \citet{tuan} for details. The two photometric techniques lead to indistinguishable results as shown in \citet{ich08}.

The irregular sampling with the large gaps makes standard Fourier transform techniques unsuitable for our data set. We therefore employ the first order structure function (SF), defined as $V(\tau) = < [s(t+\tau) - s(t)]^2 >$ with $s(t)$ being a measurement at time $t$, to look for a break in the power-law PSD \citep[a power-law in the PSD translates into a power-law in the SF albeit with a different slope, see, e.g., ][]{simonetti, hughes, tuan}.  The SF for our complete data set is shown in Fig.~\ref{sf}. The diamonds represent the high- to mid-frequency data shown in Fig.~\ref{lc}, and the crosses show the mid- to low-frequency part. The overlapping region was used to normalize the low-frequency part such that a continuous SF emerges\footnote{Equivalently, we could renormalize the high-frequency part. The reason why the SF is not automatically continuous goes back to the fact that we average the data of a whole night for the low-frequency part (which reduces the variance), whereas the high-frequency data are not averaged. Contrary to the periodogram \citep{markowitz03}, the SF does not have any normalization factor which accounts for that, and thus we use the overlapping region to determine the renormalization factor.}. It shows the superposition of a constant at short lags ($\lesssim 1\,$min) which is the result of uncorrelated measurement noise, a power-law portion between lags of minutes to hours, and a plateau/shallower power-law for long time lags ($\gtrsim 1000\,$min). This plateau at low frequencies points to a break in the power-law at shorter lags. To quantify the evidence for the existence of a break frequency and its value, Monte Carlo simulations are required, as the sampling pattern severely distorts any intrinsic power-law PSD, and introduces the effects of red-noise leak and aliasing \citep[e.g., ][]{uttley}.

\section{The simulations}

We used the following approach for our MC simulations, which is very similar to the PSRESP method by \citet{uttley} and \citet{markowitz03}: 1. An intrinsic (broken or unbroken) power-law PSD model is assumed and corresponding lightcurves are generated using the algorithm by \citet{timmer95}.  This is done independently for the high-frequency and low-frequency part. These lightcurves, which are significantly longer and more densely sampled to account for aliasing and leakage, are subsequently resampled according to the sampling function of the real data and uncorrelated measurement noise is added \citep[see][for more details]{tuan}. 2. The SF from the simulated lightcurves are calculated in the same way as is done for the data (the renormalization is done for each pair of long and short-term simulations individually). 3. An observed $\chi^2$ value is determined by using the observed SF, the mean simulated SF, and error bars equal to the rms spread of the individual simulated SFs at each time lag \citep{uttley}. 4. The goodness of fit is computed by modeling the $\chi^2$ distribution for each assumed PSD model, i.e. several thousand $\chi^2$ values are calculated using the individual realizations of the simulations instead of the observed data. The probability that the model PSD can be rejected is then given by the percentile of the simulated $\chi^2$ values exceeded by the value of observed $\chi^2$. 5. The steps above are repeated to scan a range of break frequencies and power-law slopes.

For each model PSD we used 100 simulations for each the high- and low-frequency part. By combining both sets we arrived at 10,000 simulations to determine the $\chi^2$-distribution. The class of models we employed as the intrinsic PSD is the singly broken power-law:
\begin{equation}
P(f) = \left\{ 
  \begin{array}{cc}
    A (f/f_{br})^{-\gamma}, & f \leq f_{br}, \\
    A (f/f_{br})^{-\beta}, & f > f_{br}.
  \end{array}
    \right.
\end{equation}
Here, $f_{br}$ is the break frequency (the characteristic timescale), $A$ the PSD amplitude at $f_{br}$, $\beta$ the high-frequency power-law slope, and $\gamma$ is the low-frequency power-law slope with the constraint $\gamma < \beta$. We tested a $\beta$ range of $0.5 - 3$ in increments of 0.1, a $\gamma$ range of $0 - 1$ in increments of 0.1, and a break frequency range of $10^{-6} - 3\cdot 10^{-2}\,\mbox{min}^{-1}$ in multiplicative steps of 1.2. We also tested unbroken power-law models ($P(f) \propto f^{-\beta}$) which are, however, rejected with a likelihood of acceptance of $0\%$ (i.e. none of the 10,000 simulations equalled or exceeded the observed $\chi^2$).

\section{Results \& Discussion}

\begin{figure}
\includegraphics[scale=.8]{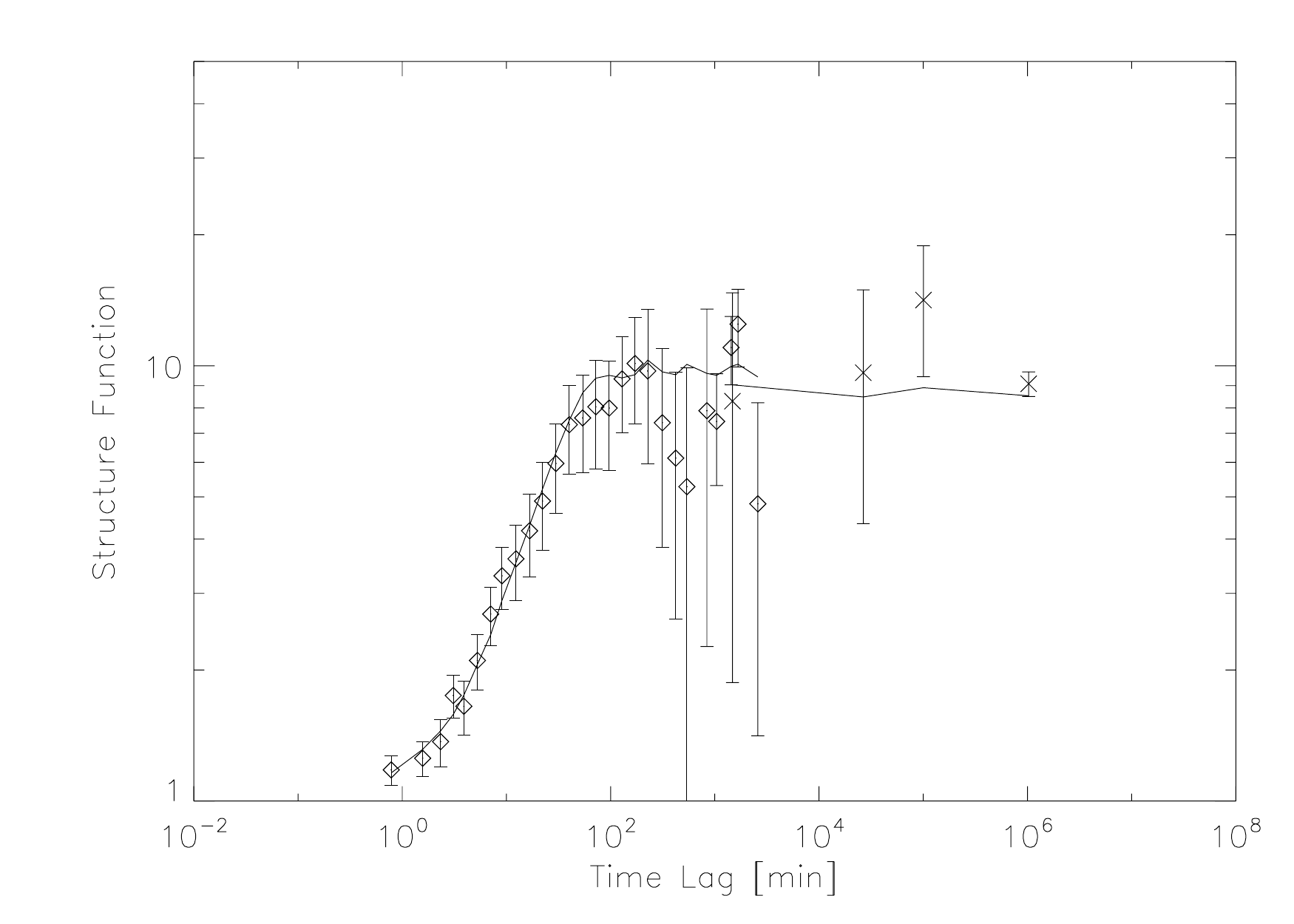}
\caption{The structure function of Sgr~A*. Diamonds represent the high- to mid-frequency data shown in Fig.~\ref{lc}, crosses show the mid- to low-frequency data (see Table~\ref{tbl1}). The solid line is our best fit with $f_{br} = 6.5\cdot 10^{-3}\, \mbox{min}^{-1}$, $\gamma = 0.3$, and $\beta = 2.1$, see Section~4. Please note that these parameters describe the PSD and cannot simply be read of the SF depicted here. Error bars are determined from our Monte Carlo simulations as described in Section~3. }
\label{sf}
\end{figure}

\begin{figure}
\includegraphics[scale=.8]{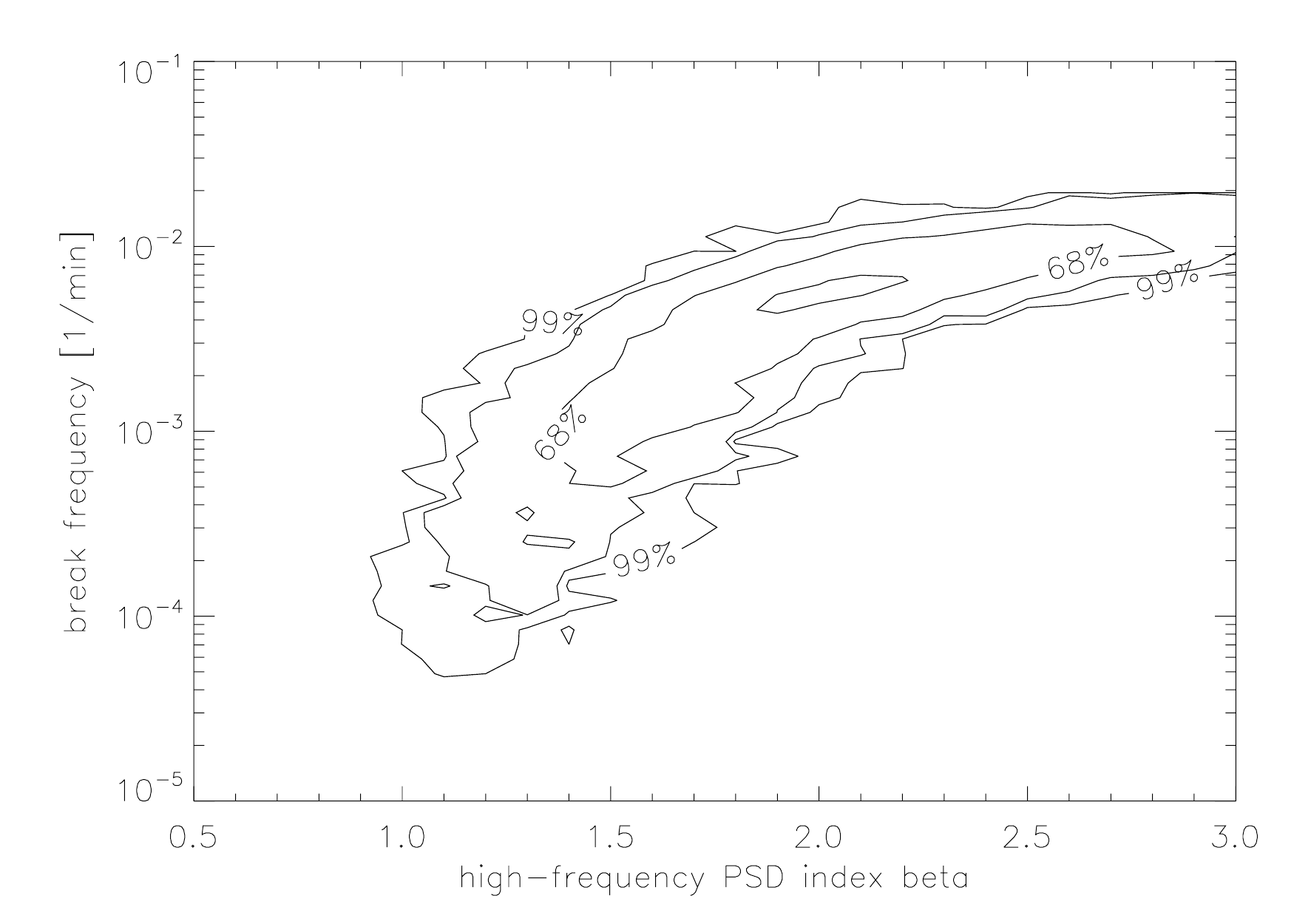}
\caption{Confidence contours showing the errors on the best fit parameters. A slice through the three dimensional parameter space at the best fitting $\gamma = 0.3$ is shown. The lines indicate the 99\%, 95\%, 68\%, and 15\% rejection probability levels.}
\label{contours}
\end{figure}

The best fit result with a likelihood of acceptance of $93\%$ can be found at $f_{br} = 6.5^{+5}_{-2.9}\cdot 10^{-3}\, \mbox{min}^{-1}$ ($\sim154^{+124}_{-87}$\,min), $\gamma = 0.3^{+0.4}_{-0.2}$, and $\beta = 2.1^{+0.5}_{-0.5}$ (errors here are 90\% confidence limits). This solution is plotted over the observed SF in Fig.~\ref{sf}. 
In Fig.~\ref{contours}  we show the confidence contours in the $\beta - f_{br}$ plane at the best fit value for $\gamma = 0.3$.  When the three dimensional distribution is marginalized over two parameters, we arrive at the following likeliest values and their 90\% confidence limits for the individual parameters: $f_{br} = 7.8^{+11.7}_{-5.6}\cdot 10^{-3}\, \mbox{min}^{-1}$ ($\sim128^{+329}_{-77}$\,min), $\gamma = 0.6^{+0.1}_{-0.4}$, and $\beta = 2.1^{+0.8}_{-0.5}$.


The existence of a break in the power-law PSD of Sgr~A* can also be inferred by a simple argument: Table 1 reveals that the mean flux of Sgr~A* during one night stays roughly the same over the 4 years of observations. If the power-law PSD extended unbroken over decades, one would expect to observe very different mean fluxes over a time baseline of years\footnote{The high-frequency power-law slope of Sgr~A* is close to 2. In this simple case of Brownian motion -- assuming there is no break in the PSD -- the intrinsic standard deviation of the flux scales with the square root of time.}. Our detailed MC simulations serve the purpose of quantifying the location of the power-law break. The same simple argument also holds true for the mean X-ray flux of Sgr~A*. 

Our result of a break frequency at $\sim 6.5\cdot 10^{-3}\, \mbox{min}^{-1}$ is strikingly different from the prediction of the scaling relation proposed by \citet{mchardy}. Using a mass of $M=4\cdot 10^6\, M_\sun $ \citep{ghez08} and a bolometric luminosity of $2\cdot 10^{36}$ erg/s \citep{narayan}, the relation by \citet{mchardy} predicts a break frequency at the order of $10^{-9}\, \mbox{min}^{-1}$ for Sgr~A*. 

\begin{figure}
\includegraphics[scale=.6]{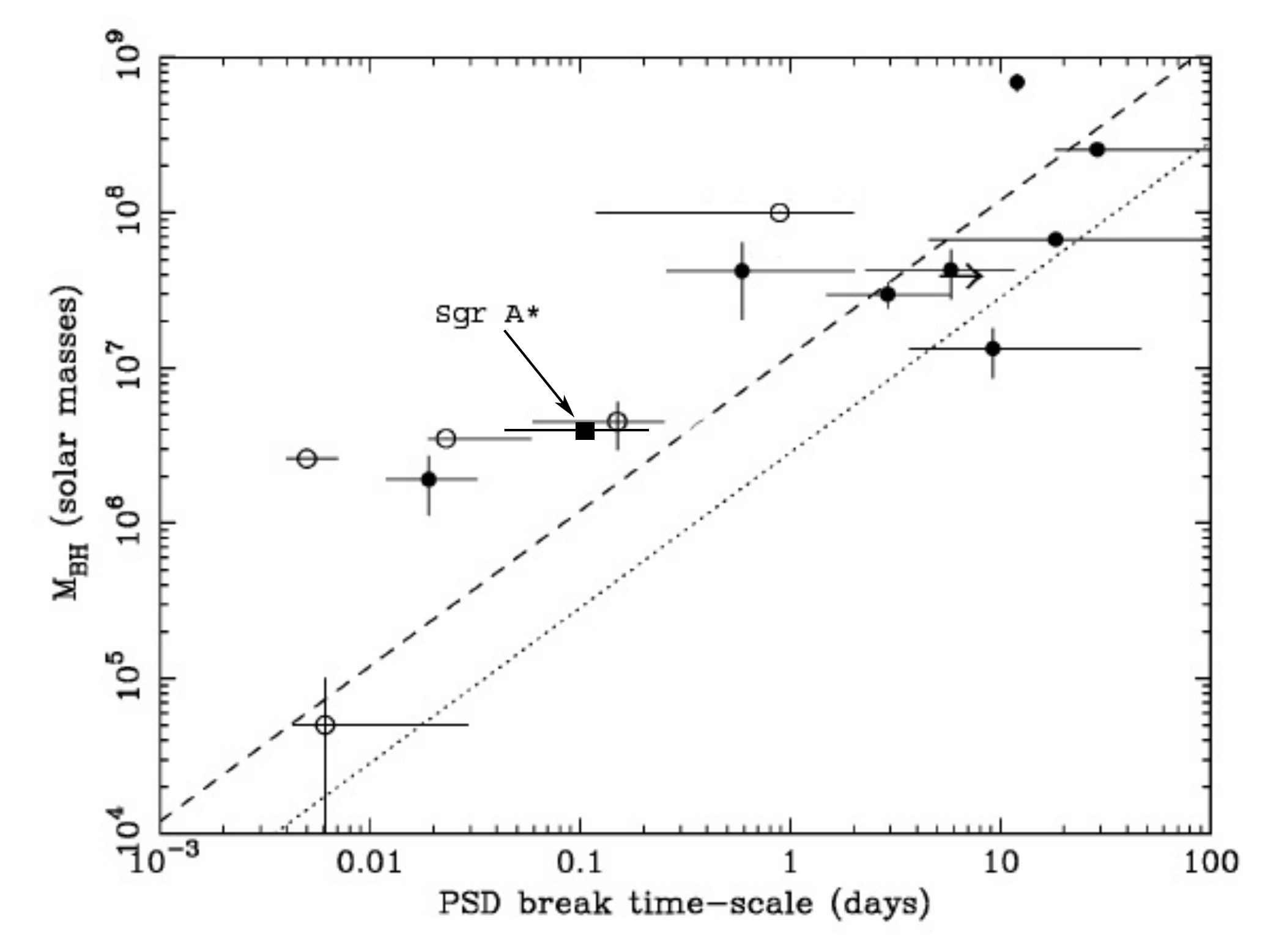}
\caption{Sgr~A*'s break timescale as reported in this paper over plotted onto Figure~11 of \citet{uttley05}, which shows BH mass versus PSD break timescale for various AGN. The mass of Sgr~A* has been taken from \citet{ghez08}. Its uncertainty corresponds to the height of the black square. Filled circles mark masses determined from optical reverberation mapping, open circles represent masses determined using other methods. The straight lines represent the expected relations if linear mass scaling is assumed from the typical timescales observed in the high/soft (dashed line) and low/hard (dotted line) state of the BHXRB Cyg X-1 (assuming $10\,\mbox{M}_\sun$ for its mass). 
}
\label{f4}
\end{figure}

Interestingly, if the term with the bolometric luminosity is neglected in the \citet{mchardy} relation, i.e. the bolometric index (called 'B' in their paper) is set to zero, their relation predicts a break timescale of $\sim 110$\,min for Sgr~A*. Also, our deduced break timescale fits the expected value if linear scaling with BH mass is assumed from the typical break timescales observed in the high/soft and low/hard states of Cyg X-1; see Fig.~\ref{f4}. This  suggests that, while the inclusion of the bolometric luminosity as a free parameter improves the fit in a limited luminosity range, it is in fact not the true physical correction factor that can explain the scatter around the linear mass scaling. 

It is, however, uncertain how meaningful a comparison of  Sgr~A*'s break timescale with the breaks of the much higher luminosity AGN used to derive the mass-luminosity-timescale relation of \citet{mchardy}  is. These Seyferts are almost certainly in a different accretion state. They have a prominent optically thick accretion disk  and much higher accretion rates. The \citet{mchardy} relation seems to work  well when extrapolated to  soft state  BHXRBs, which are at comparable Eddington accretion rates to the Seyferts, but for the low-accretion rate hard states the comparison is less obvious. In fact, work by \citet{gierlinski} suggests that in the hard state the high-frequency shape of the PSD is  constant despite large changes in luminosity. Therefore, we conclude that the break in the NIR PSD of Sgr~A* implies  that the \citet{mchardy} relation with luminosity scaling applies to soft state AGN, but that in the hard state there is only mass scaling. Then Sgr~A* seems to fit in well with the hard state BHXRBs, consistent with its low luminosity.


Future work will further elucidate the relationship between variability processes across the range of BH masses. Our work presented here extends the set of supermassive BHs with determined break frequencies. The BH in our Galactic center is especially important, as it is the most underluminous source (in terms of Eddington luminosity) accessible to observations, and its mass determination is the most precise for supermassive BHs \citep{ghez08,gillessen08}.

\acknowledgements
We want to thank Phil Uttley and Matt Malkan for their constructive comments on the manuscript. LM is supported by a fellowship within the Postdoc-Program of the German Academic Exchange Service (DAAD). Some of the data presented here were obtained from Mauna Kea observatories. We are grateful to the Hawai'ian people for permitting us to study the universe from this sacred summit. This work was supported by NSF grant AST-0406816.

{\it Facilities:} \facility{VLT:Yepun (NACO)}, \facility{Keck:II (NIRC2) }

\clearpage

\begin{deluxetable}{cccccc}
\tabletypesize{\scriptsize}
\tablecaption{Summary of the data \label{tbl1}}
\tablewidth{0pt}
\tablehead{
\colhead{Date} & \colhead{Telescope\tablenotemark{a}} & \colhead{Filter} & \colhead{Duration} & \colhead{Mean Flux\tablenotemark{b}} &
\colhead{published in} \\
\colhead{(UT)} & \colhead{} & \colhead{} & \colhead{(min)} & \colhead{(mJy, de-reddened)} &
\colhead{} 
}
\startdata
2004 July 06 & Keck & L' & 165 & 2.64    &  Hornstein 2007 \\
2004 July 07 &  Keck/VLT & L'/K & 532 & 4.05 & Hornstein 2007/Eckart+ 2006a\\
2004 July 08 & VLT & K & 360 & 2.00 & Eckart+ 2006a\\
2004 July 26 & Keck & K & 42 & 4.43 & Ghez+ 2008\\
2005 July 31 & VLT/Keck & K & 591 & 4.01 & Meyer+ 2008\\
2006 May 03 & Keck & K & 140 & 5.53 & Do+ 2008\\
2006 June 20 & Keck & K & 125 & 4.57 & Do+ 2008\\
2006 June 21 & Keck & K & 164 & 3.62 & Do+ 2008\\
2006 July 17 & Keck & K & 189 & 2.86 & Do+ 2008\\
2007 May 18 & Keck & K & 84 & 5.53 & Do+ 2008\\
2007 August 12 & Keck & K & 57 & 3.05 & Do+ 2008\\
2008 May 15 & Keck & K & 153 & 4.57 & unpublished\\
2008 June 02 & Keck & K & 160 & 4.00 & unpublished\\
2008 July 24 & Keck & K & 179 & 3.51 & unpublished
\enddata
\tablecomments{The first three entries are used for the high- to mid-frequency coverage, see Fig.~\ref{lc}.}
\tablenotetext{a}{Keck means NIRC2 at Keck II; VLT means NACO at VLT (Yepun).}
\tablenotetext{b}{The fluxes are K-band fluxes. When the L' filter was used, we scaled the flux down to the K-band level using a spectral index of $\alpha = -0.6$ \citep{hornstein07}.}
\end{deluxetable}

\clearpage

\end{document}